\documentclass[12pt, twoside]{article}
\usepackage{a4wide,amssymb,amsmath}
\usepackage{units}
\usepackage{slashed}
\usepackage{hyperref}

\newcommand{\be}{\begin{equation}}
\newcommand{\ee}{\end{equation}}
\newcommand{\bea}{\begin{eqnarray}}
\newcommand{\eea}{\end{eqnarray}}

\newcommand{\cL}{\mathcal{L}}
\newcommand{\cM}{\mathcal{M}}
\newcommand{\diag}{\mathop\mathrm{diag}}
\newcommand{\tr}{\mathop\mathrm{tr}}
\newcommand{\abp}{|\mathbf{p}|}
\newcommand{\abpp}{|\mathbf{p}'|}
\newcommand{\ct}{\cos\theta}

\begin{document}

\begin{titlepage}

\rightline{CERN-PH-TH/2011-293}

\begin{centering}
\vspace{1cm}
{\large {\bf Model dependence of the bremsstrahlung effects  \\ 
\vspace{0.1cm} from the superluminal neutrino at OPERA}}\\

\vspace{1.5cm}

{\bf Fedor Bezrukov}$^{a,*}$ and {\bf Hyun Min Lee}$^{b,\dagger}$
\\
\vspace{.2in}

$^{a}${\it Arnold Sommerfeld Center for Theoretical Physics,
  Department f\"ur Physik,
  Ludwig-Maximilians-Universit\"{a}t, Theresienst.~37, 80333
  M\"{u}nchen, Germany} 

\vspace{.1in}

$^{a}${\it Institute for Nuclear Research of the Russian Academy of Sciences, 60th October Anniversary prospect 7a, Moscow 117312, Russia.}

\vspace{.1in}

$^{b}${\it CERN, Theory division, CH-1211 Geneva 23, Switzerland.} 

\end{centering}
\vspace{2cm}

\begin{abstract}
\noindent
We revisit the bremsstrahlung process of a superluminal neutrino motivated by OPERA results. 
From a careful analysis of the plane wave solutions of the superluminal neutrino, we find that the squared matrix elements contain additional terms from Lorentz violation due to the modified spin sum for the neutrino. We point out that the coefficients of the decay rate and the energy loss rate significantly depend on the details of the model, although the results are parametrically similar to the ones obtained by Cohen and Glashow \cite{cohenglashow}.
We illustrate this from the modified neutral current interaction of neutrino with Lorentz violation of the same order as in the modified dispersion relation.

\end{abstract}

\vspace{5 cm}

\begin{flushleft}

$^{*}$ E-mail address: \nolinkurl{Fedor.Bezrukov@physik.uni-muenchen.de} \\
$^{\dagger}$ E-mail address: \nolinkurl{Hyun.Min.Lee@cern.ch}

\end{flushleft}

\end{titlepage}

\section{Introduction}

Lorentz invariance is one of the cornerstones of the modern quantum
field theory, and it was completely compatible with all previous
experiments and observations.  Recently, an intriguing result has been
presented by the OPERA collaboration \cite{opera}, claiming that the
muon neutrino speed exceeds the speed of light by $(v-c)/c\simeq
2.37\pm0.32^{+0.34}_{-0.24}\times10^{-5}$.  This result has been recently confirmed by a test
performed using short-bunch wide-spacing beam in the revised version
of Ref.~\cite{opera}. There are similar measurements from other
oscillation experiments, for example, MINOS, resulting in
$(v-c)/c=5.1\pm2.9\times10^{-5}$ \cite{MINOS}, but the OPERA is the
first experiment which observed a positive $(v-c)$ for neutrinos with
high statistical significance of about $6.2\sigma$.  There is no
statistically significant energy dependence of superluminality at
OPERA where the results for the low- and high-energy samples with
averaged neutrino energies, $13.8$ GeV and $40.7$ GeV, respectively,
are in agreement \cite{opera}.

Such a drastic result is seemingly in contradiction with a set of
other neutrino observations \footnote {A review on the bounds on Lorentz violation in the neutrino sector before OPERA can be found in Ref.~\cite{preOPERA}.}. The detection of neutrinos emitted from
the SN1987A supernova \cite{SN1987A1,SN1987A2,SN1987A3} puts a stringent bound on the
electron anti-neutrino speed, $|v-c|/c<2\times10^{-9}$.  Furthermore,
the observation of neutrino oscillations demands that neutrino
velocities for different neutrino flavors should be equal up to
$|v_i-v_j|\lesssim10^{-19}$ for $i\neq j$, otherwise the coherence is lost
and the oscillation pattern is smeared
\cite{CGdispersion,neutrino_osc}.  
This contradiction could be
explained in two ways.  One is to make the neutrino velocity energy
dependent \cite{giudice,ellis}, because the typical neutrino energy
is \unit[10]{MeV} for the supernova neutrinos, and is about
\unit[28]{GeV} for the CNGS neutrino beam, used by OPERA.  Another is
to make the neutrino superluminal only within the Earth radius
\cite{DvaliVikman,ellis,allothers1,allothers2,allothers3}, or only inside matter
\cite{medium}. A different route to the solution could be taken by considering models with energy non-conservation or deformed Lorentz invariance \cite{deformLV} in the neutrino interactions. But this would be hard to formulate in the language of ordinary quantum field theory so we do not pursue this option in this article.

In any model explaining the OPERA results it is important to check,
whether the creation, propagation, and detection in the OPERA setup
can be explained.  There are very strong statements that invalidate
most of the proposed models for the OPERA results.  First, the
superluminal neutrino can radiate electron-positron pairs (in a way
analogous to the Cherenkov radiation) \cite{cohenglashow}, thus
loosing energy before reaching the detector.  Second, the decay of a
fast moving pion is modified, and even the initial neutrino spectrum
should have a strong cutoff at energy, which is below the average
energy detected by OPERA \cite{piondecay1,piondecay2,piondecay3}.  Both of these results rely
on the following assumption---the only thing modified in the theory is
the dispersion relation of the neutrino.  As the neutrino speed is
given by the derivative of the dispersion relation $v=dE/dp$, a
constant neutrino speed at OPERA means that the neutrino dispersion
relation has the form $E=vp\equiv(1+\delta/2)p$ with
$\delta\simeq5\times10^{-5}$.  A stronger claim, based on the result
of Ref.~\cite{cohenglashow} was made by the analysis of the ICARUS
results \cite{icarus}.  The ICARUS detector, while not being able to
measure the arrival time of the neutrinos, can carefully measure their
energy spectrum.  The comparison of the expected and measured spectra
provides strict bounds on the neutrino speed, because the energy loss
by electron-positron radiation \cite{cohenglashow} would significantly
change the spectrum.

In this article we re-analyze the bremsstrahlung process of a superluminal neutrino that has been done in Ref.~\cite{cohenglashow}.
The main point of the previous analysis follows from the kinematical possibility for the neutrino with non-standard dispersion relation to ``decay'' into other
particles.  For example, for the process $\nu\rightarrow \nu+e^++e^-$, the
``masses'' (or squares of the four momenta) of the initial and final
neutrinos are different, enabling the process to take place.  However, the exact
calculation for the process rate is more involved.  
Specifically, when the dispersion relation for neutrino is modified, one has to use the modified plane wave solutions for the neutrino. In turn, the spin sum for the final neutrino gets modified, leading to the additional terms of the order of $\delta$ in the expression for the squared matrix elements, as compared to the previous calculations ~\cite{cohenglashow}.  
Moreover, when there are modifications of the same order in
the electroweak interaction vertex of neutrino as in the neutrino dispersion relation, there appear more
terms of the similar order in the squared matrix elements too. 
All these effects are added up to give a nontrivial result, which depends on the additional modifications of the same order.  
This is due to the fact that the squared matrix elements (obtained
by the standard rules for the spin sum for Lorentz invariant fermions) is only of the order 
of $\delta^2$ in the kinematically allowed region.
As a result, due to various cancellations in the matrix element, the
final probability of the bremsstrahlung process depends on the details
of the Lorentz violation in the model.

We explicitly construct (at the level of Fermi four fermion 
interactions) two models with broken Lorentz symmetry.  Both models have a common property that the neutrino kinetic term contains a Lorentz violating term
as inspired by the modification of the metric for the neutrino
\cite{DvaliVikman}.  The difference is that one of the models keeps the interaction terms Lorentz invariant
while the other model introduces a similar Lorentz violation in
the electroweak neutral current of neutrino too.

In section \ref{sec:model} we introduce the Lagrangians for the
models. Then, in section \ref{sec:spinsum} we obtain the free solutions for the
neutrinos and present the rules for ``summation'' over the spin
states.  Consequently, in section \ref{sec:width}  we provide the detailed
calculation of the neutrino decay width and the rate of energy loss.
Finally, conclusions are drawn.

\section{Models}
\label{sec:model}

We first define the framework for the calculation of the decay
(bremsstrahlung) of a superluminal neutrino into an electron-positron
pair.  Following the ideas in \cite{CGdispersion,kostelecky} the lowest order
Lorentz violating operator in the Lagrangian for the massless (Weyl)
fermion looks like
\begin{equation}
  \label{eq:1}
  \cL = i\bar\nu \gamma^\mu\tilde{g}_{\mu\nu}\partial^\nu
                 (1-\gamma^5) \nu,
\end{equation}
where the Lorentz violating ``metric'' can be chosen as
\begin{equation}
  \label{eq:2}
  \tilde{g}_{\mu\nu} = \diag( 1, -v, -v, -v ),
\end{equation}
with the neutrino speed $v\equiv1+\delta/2$.  Quite obviously, this
action gives rise to the superluminal neutrino dispersion relation,
$E=v|\mathbf{p}|$. In the limit of Lorentz invariance, we get ${\tilde g}_{\mu
\nu}=\diag (1,-1,-1,-1)\equiv g_{\mu\nu}$.

The next component of the model is the interaction term.  We will not
go in the details of how the model emerges from the underlying electroweak gauge theory, because this would lead to complications due to the different velocities for the left-handed electron and neutrino of the same multiplet.
We will take the purely phenomenological approach and analyze two types of the four
fermion neutral current interaction.  The first one will be the usual
Lorentz invariant one (model I)
\begin{equation}
  \label{muonnu}
  \cL_{\mathrm{int1}} =
  \frac{G_F}{\sqrt{2}}\Big[{\bar \nu}
  \gamma_\mu(1-\gamma_5)\nu\Big]\Big[{\bar
    e}\gamma^\mu(v_e-a_e\gamma^5)e \Big].
\end{equation}
The second one is inspired by a ``gauge invariant'' Lagrangian, where
the covariant derivative enters in the same way as in (\ref{eq:1})
(model II)
\begin{equation}
  \label{eq:4}
  \cL_{\mathrm{int2}} =
  \frac{G_F}{\sqrt{2}}\Big[{\bar \nu}
  \gamma^\mu(1-\gamma_5)\nu\Big]
  \tilde{g}_{\mu\nu}
  \Big[{\bar
    e}\gamma^\nu(v_e-a_e\gamma^5)e \Big].
\end{equation}
In fact, this interaction term does not follow from the
gauge invariant SM, as far as electroweak gauge symmetry is unbroken.  
In this work, we assume this possibility but do not consider a
microscopic model with Lorentz
symmetry/electroweak symmetry breaking for that.

\section{Free solutions for the neutrinos and spin sums}
\label{sec:spinsum}

The action (\ref{eq:1}) leads to the Dirac equation of the form (in
momentum representation and two component form for simplicity)
\begin{equation}
  \label{eq:3}
  (E \sigma^0 -v p^i \sigma^i) \chi = 0,
\end{equation}
where $\sigma^0$, $\sigma^i$ are the unit $2\times2$ matrix and Pauli matrices, respectively.  Let us also (using
invariance under O(3) spatial rotations) align the momentum along the
3rd spatial axis.  Then we immediately get two solutions
\begin{eqnarray}
  \label{eq:5}
  E=vp^3  &\text{ with }& \chi=\sqrt{2E}\begin{pmatrix} 1 \\ 0 \end{pmatrix}, \\
  E=-vp^3 &\text{ with }& \chi=\sqrt{2E}\begin{pmatrix} 0 \\ 1 \end{pmatrix},
\end{eqnarray}
where $\sqrt{2E}$ is the standard overall normalization.
The first solution corresponds to the neutrino and the second solution to the antineutrino, both with
velocity $v$.  Notice that the dispersion relation is modified, while
the spinors have the usual form.  The spin ``sum'' for the neutrino (which is in this
case trivial, as far as there is only one spin state for the neutrino)
is
\begin{equation}
  \label{eq:6}
  \sum_{s=1/2} \chi_s \chi_s^\dagger
  = 2E\begin{pmatrix} 1 & 0 \\ 0 & 0 \end{pmatrix}
  = p^\mu\tilde{g}_{\mu\nu}{\bar\sigma}^\mu
  \neq p^\mu g_{\mu\nu}{\bar \sigma}^\mu.
\end{equation}
The important observation here is that the spin sum is not given by
the standard expression, but with the momentum contracted with the
sigma matrices using the \emph{superluminal} metric.  The difference is
nontrivial in the first order in $\delta$, which is essential for the
calculation of the decay width.

The generalization to the usual four component spinors is obvious,
\begin{equation}
  \label{eq:7}
  \sum_s \nu_s \bar{\nu}_s = p^\mu \tilde{g}_{\mu\nu} \gamma^\nu
  \equiv \tilde{\slashed{p}},
\end{equation}
where the momentum with the tilde is a shorthand for
$\tilde{p}_\mu\equiv\tilde{g}_{\mu\nu}p^\nu$.  Note, that raising and
lowering the indices is always done with the normal metric, while the tilde
means the additional factor of $v$ for the spatial part in the scalar product.

Now we are ready to evaluate the matrix elements and the squared of them.

\section{Decay width calculation}
\label{sec:width}

Following the standard rules for the decay process, we get for the bremsstrahlung process
$\nu_\mu\to\nu_\mu+e^++e^-$ with four momenta $p$, $p'$, $q_1$, and
$q_2$,
\begin{equation}
  \label{eq:8}
  \Gamma = \frac{(2\pi)^4}{2E_\nu}\int
  \prod_f \frac{d^3{\vec p}_f}{(2\pi)^3 2E_f}\delta^4(p-p'-q_1-q_2)
  \cdot \sum_{\mathrm{spin}}|{\cM}|^2.
\end{equation}
Here, note that we do not average over the spin states of the
neutrino, which was done in the calculation of \cite{cohenglashow}.
In the current setup the neutrino has explicitly only one spin state, instead of two spin states for a massive neutrino.
The square of the matrix
elements is the following: in the model I (\ref{muonnu}),
\begin{equation}
  \label{eq:9}
 \sum|\cM_\mathrm{I}|^2 = \frac{G_F^2}{2} M^{\alpha\beta}E_{\alpha\beta}
  ,
\end{equation}
and in the model II (\ref{eq:4}),
\begin{equation}
  \label{eq:10}
  \sum|\cM_\mathrm{II}|^2 = \frac{G_F^2}{2}
 M^{\alpha\beta}\tilde{g}_{\alpha\gamma}\tilde{g}_{\beta\delta}E^{\gamma\delta}
  .
\end{equation}
Here the individual traces are
\begin{align}
  M^{\alpha\beta} &=
  \tr\Big[
  \tilde{\slashed{p}}' \gamma^\alpha(1-\gamma_5) \tilde{\slashed{p}} (1+\gamma_5)\gamma^\beta
  \Big], \\
  E_{\alpha\beta} &=
  \tr\Big[
  \slashed{q}_2 \gamma_\alpha(v_e-a_e\gamma_5) \slashed{q}_1 (v_e+a_e\gamma_5)\gamma_\beta
  \Big].
\end{align}
There are two differences from the standard calculation with Lorentz invariance: the spatial
parts of $p$ and $p'$ momenta in $M^{\alpha\beta}$ are multiplied with
$v$, and the indices between $M$ and $E$ are contracted with
superluminal metric for the model II. 

The results of the multiplication are (in the approximation of a
purely axial electron neutral current, $a_e=-1/2$ and $v_e=0$)
\begin{equation}
 \sum|\cM_\mathrm{I}|^2= 8G_F^2\Big[
  (\tilde{p} q_1)(\tilde{p}'q_2) + 
  (\tilde{p}' q_1)(\tilde{p} q_2) \Big], \label{msmodel1}
\end{equation}
and
\bea
  \label{eq:11}
  \sum|\cM_\mathrm{II}|^2&=& 8G_F^2\Big[
  (\tilde{p} \tilde{q}_1)(\tilde{p}'\tilde{q}_2) + 
  (\tilde{p}' \tilde{q}_1)(\tilde{p} \tilde{q}_2) \nonumber \\
  &&\quad \quad \quad -(\tilde{p}\tilde{p}')(\tilde{q}_1\tilde{q}_2)
  -(\tilde{\tilde{p}}\tilde{\tilde{p}}')(q_1q_2)
  +\frac{1+3v^2}{2}(\tilde{p}\tilde{p}')(q_1q_2)
  \Big].
\eea
Here the tilde always means one factor of $v$ in front of the spatial
product, i.e.~$({\tilde p}q)\equiv p^0 q^0-v\mathbf{p\cdot q}$, $(\tilde{p}\tilde{q})\equiv p^0q^0-v^2\mathbf{p\cdot q}$,
$(\tilde{\tilde{p}}\tilde{\tilde{p}}')\equiv p^0p'^0-v^4\mathbf{p\cdot p'}$,
etc.  Note that in model II the second line in (\ref{eq:11}) does not vanish. For comparison to our model I, in Ref.~\cite{lietal}, the second term in the squared amplitude (\ref{msmodel1}) was missing, the spin average for the initial neutrino was taken, and the modified plane-wave solutions for neutrino were not taken into account for the spin sum of the neutrino.

The rest of the calculation is rather straightforward, and consists
just of careful integration over the final momenta.  The safest way is
to perform the calculation in the lab frame directly.  We will
only sketch the derivation here.

First, we perform integration over the momenta of the electron and
positron, in the limit of zero electron mass (this is fine as far as we
are interested in the decay of high energy neutrinos)
\begin{equation}
  \int q_{1\mu}q_{2\nu}
  \frac{d^3\mathbf{q}_1}{E_1}\frac{d^3\mathbf{q}_2}{E_2}
  \delta^4(k-q_1-q_2)
  = \frac{\pi}{6}(k^2 g_{\mu\nu}+2k_\mu k_\nu ),
\end{equation}
where $k\equiv q_1+q_2$ is the momentum of the electron-positron pair.
It is convenient to rewrite the remaining integration as the integration over
the modulus of the final neutrino momentum $\abpp$ and the cosine of the
angle $\theta$ between $\mathbf{p}$ and $\mathbf{p'}$ (in the lab
frame).  Calculating all the scalar products together with the dispersion relations,
$p^0=v\abp$ and $p^{\prime 0}=v\abpp$, we obtain the decay rate as follows,
\be
\Gamma=\frac{G^2_F}{96\pi^3v^2\abp}\int \abpp d\abpp d\ct \,\, {\cal I}
\ee
where ${\cal I}={\cal I}_{\mathrm {I,II}}$ for models I and II are given by
\begin{multline}
  \label{eq:16}
{\cal I}_{\mathrm {I}} =\Big( (v^2 - 1) (\abp^2 + \abpp^2) - 2 \abp \abpp (v^2 - \ct)\Big) v^2 \abp \abpp (1 - \ct)  \\ - 
   2 v^2 \Big(  (v - 1) \abp^2 - 
      \abp \abpp (v - \ct) \Big) \Big((v - 1) \abpp^2 - 
      \abp \abpp (v - \ct) \Big)
      \end{multline}
and
\begin{multline}
  \label{eq:15}
{\cal I}_{\mathrm {II}}=  \bigg[(1 + 3 v^2) \Big((v^2 - 1) (\abp^2 + \abpp^2) - 2 \abp \abpp (v^2 - \ct) \Big) + 
     2 v^2 \abp \abpp (1 - \ct)\bigg]  v^2 \abp \abpp (1 - \ct) \\- 
  2 \Big( (v^2 - 1) (\abp^2 + \abpp^2) - 2 \abp \abpp (v^2 - \ct)\Big) v^2 \abp \abpp (1 - v^2 \ct) - 
  2 v^4  \abp^2 \abpp^2 (1 - \ct)^2.
\end{multline}
The integration over momenta is governed by the
positivity of the electron-positron pair invariant mass, $k^2>0$.  For $v=1+\delta/2$ this
means that  we should
integrate over the whole region $-1<\ct<1$ for $0<\abpp<p_c\equiv \abp\delta/(4+\delta)$, but only over $1>\ct>\cos\theta_\mathrm{min}\simeq(-\delta \abp^2-\delta
\abpp^2+2(1+\delta)\abp \abpp)/(2\abp \abpp)$ for $p_c<\abpp<\abp$.  The latter region gives in fact
the major contribution to the integral.
The integral for the rate of the energy loss is similar but with
the integrand multiplied by $-(E-E')$.

Performing the momentum integrals we obtain the decay rate and the rate of the energy loss as follows,
\begin{align}
  \label{eq:12}
  \Gamma &=a \frac{G_F^2}{192\pi^3} E^5,\\
  \frac{dE}{dx} &= - a' \frac{G_F^2}{192\pi^3} E^6
\end{align}
Although the results are parametrically similar to the ones in Ref.~\cite{cohenglashow}, the numerical coefficients turn out to be model dependent.
We have that for the model I, 
\bea
  \label{eq:13}
  a_\mathrm{I} &=& \frac{1}{420}(v-1)^3(v+1)(53+20v-5v^2)\simeq\frac{17}{420}\delta^3,  \\
  a'_\mathrm{I} &=& \frac{1}{672}(v-1)^3(v+1)v(67+28v-7v^2)\simeq \frac{11}{336}\delta^3,
\eea
and for the model II,
\bea
\label{eq:14}
 a_\mathrm{II} &=& \frac{1}{420}(v^2-1)^3(5v^2+19)\simeq 
  \frac{2}{35}\delta^3,
  \\
  a'_\mathrm{II} &=& \frac{1}{672}(v^2-1)^3v(7v^2+23)\simeq \frac{5}{112}\delta^3.
\eea
For comparison, the results in Ref.~\cite{cohenglashow}, which are obtained with
the standard ``Lorentz invariant'' expression for the squared matrix element, are
\bea
 a_\mathrm{CG} &=&\frac{(v^2-1)^3}{14v^2}\simeq \frac{1}{14}\delta^3,
  \\
  a'_\mathrm{CG} &=& \frac{25}{448}\frac{(v^2-1)^3}{v}\simeq\frac{25}{448}\delta^3.
\eea
We find that in all the cases, the decay rate and energy loss are proportional to $\delta^3$. This $\delta^3$ dependence in our models reminds us of the argument in Ref.~\cite{cohenglashow} based on the kinematics that a superluminal neutrino gets an ``effective'' mass such that the decay rate is proportional to $\delta^3 E^5$.
However, since the numerical coefficients are model dependent, it might be possible to construct models of Lorentz violation that allow for a reduction or cancellation of $\delta^3$ terms. 

We obtain the lifetime of a superluminal neutrino as compared to the results in  \cite{cohenglashow}: $\tau=\Gamma^{-1}$ is $1.76(1.25)\,\tau_\mathrm{CG}$ for the model I(II).
The mean fractional energy loss due to a single pair emission is
$E^{-1}(dE/dx)/\Gamma\simeq 0.81(0.78)$ for model I(II), which is very similar to $0.78$, the value given in \cite{cohenglashow}.
The terminal energy of the superluminal neutrino is given by \cite{cohenglashow}
\begin{equation}
  E_T = \Big(\frac{5G_F^2}{192\pi^3}\,a' L\Big)^{-1/5}.
\end{equation}
For the OPERA baseline of $\unit[730]{km}$, we have $\unit[13.9]{GeV}$ and
$\unit[13.1]{GeV}$ for the models I and II, which is numerically very
close to the value $\unit[12.5]{GeV}$ in \cite{cohenglashow}.

\section{Conclusions}

We calculated the decay rate and the energy loss of a superluminal
neutrino in two models where the Lorenz violation is introduced in
the kinetic term of the neutrino in the action.  We found that due to the
change in the form of the solutions of the free field (Weyl) equation
for neutrinos, the modified spin sum rules must be used for the calculation of the
matrix element.  We also found that the final result depends explicitly on
the form of the Lorentz violation in the action.  In the analyzed
models, the energy loss by electron-positron bremsstrahlung still makes
the models incompatible with the observation of high energy neutrinos
at OPERA (and ICARUS as well as IceCube \cite{cohenglashow,boundCG}), but it advises us for a very careful calculation of
the neutrino decay rate in more complicated models (for example,
models with energy dependent velocity or modified velocities of
electrons).  A model with cancellation of $\delta^3$ contribution in
the decay rate (if it exists) may evade the neutrino decay constraint.

\section*{Acknowledgments}

FB is grateful to Dmitry Levkov for discussions about the basic
properties of the quantum field theory. HML thanks Gian Giudice for the discussion on superluminal neutrinos. The work of FB is partially supported
by the Humboldt foundation.  HML is partially supported by Korean-CERN
fellowship.


\end{document}